\documentclass[prb,amsmath,amssymb,twocolumn,superscriptaddress]{revtex4-1}

\usepackage[dvipdfmx]{graphicx}
\usepackage{dcolumn}
\usepackage{bm}
\usepackage{color}

\usepackage{amsmath}	
\usepackage{amsfonts}
\usepackage{braket}
\usepackage{amsthm}

\newtheorem*{claim}{Claim}

\begin{document}

\title{
Quantized polarization in a generalized Rice-Mele model at arbitrary filling
}

\author{Yasuhiro Tada}
\email[]{ytada@hiroshima-u.ac.jp}
\affiliation{
Quantum Matter Program, Graduate School of Advanced Science and Engineering, Hiroshima University,
Higashihiroshima, Hiroshima 739-8530, Japan}
\affiliation{Institute for Solid State Physics, University of Tokyo, Kashiwa 277-8581, Japan}

\begin{abstract}
We discuss the charge polarization in a generalized Rice-Mele model at arbitrary particle filling per site as a model 
of charge ordered systems in one dimension.
The model possesses neither the conventional bond-centered inversion symmetry nor the one site
translation symmetry alone, but has combinations of these symmetries.
We show that the charge polarization in the ground state is quantized by the combined symmetry
and is characterized solely by the filling.
Especially, the polarization can be $1/2$ (mod 1) in the zero filling limit.
Under the open boundary condition, there exist excess charges accumulated near edges of the system
irrespective of existence or absence of edge modes.
Correspondingly, we decompose the polarization into 
a bulk contribution and an edge contribution,
and numerically demonstrate that the polarization is dominated by the former (latter) 
when the energy gap is small (large).
We also discuss a simple generalization of our model and examine absence/existence of a
gapless edge mode protected by the inversion symmetry
by introducing intra unit cell and inter unit cell contributions of the
charge polarization.
\end{abstract}

\maketitle

\section{introduction}

An insulator can have polarization which characterizes asymmetry of a charge configuration.
A classical example is the ionic insulator where positive point charges and 
negative point charges are alternatively distributed.
Careful discussions are necessary for an appropriate definition of the polarization even for such a simple system
under the periodic boundary condition.
It is now widely accepted that the polarization can well be described by the Berry-Zak formula,
which clarifies an intimate relationship of the polarization to topology and symmetry
~\cite{Resta2007,Vanderbilt2018,Spaldin2012,KingSmith1993,Vanderbilt1993,Zak1989}.
The celebrated prototypical model of dipole topological insulators is the Su-Schrieffer-Heeger (SSH) model,
where the hopping integral has a staggered modulation~\cite{SSH1979,SS1981,SSHbook}.
The polarization of the SSH model is quantized by an inversion symmetry,
and 
there have been extensive theoretical studies of the generalized SSH models and many interesting properties
have been clarified~\cite{Li2014,Yuan2017,Perez2019,Scollon2020,Guo2015,Alvarez2019,Huda2020,Maffei2018,Xie2019,
Hetenyi2021}.
Topological nature of the charge polarization is clearly seen in charge pumping where 
a change of the polarization in an adiabatic cycle in a parameter space is quantized by the 
Chern number~\cite{Thouless1983,ThoulessPump2023}.
Indeed, the Rice-Mele (RM) model~\cite{RM1982} was experimentally realized in cold atoms and 
the Thouless pumping was observed~\cite{Lohse2016,Nakajima2016,Nakajima2021}.
In this way, the SSH and RM models are regarded as one of the most important models in the context of topological insulators.

The very basic physics behind the SSH and RM models is charge ordering due to the Peierls instability
~\cite{Kagoshima1981,Smaalen2004}
as they were originally proposed for polyacetylene and diatomic polymers~\cite{SSH1979,SS1981,RM1982}.
A trimerized variant of the SSH model was also studied with an application to the Peierls system
TTF-TCNQ~\cite{SS1981}, although related models have been discussed in relation to various systems
such as engineered materials and cold atoms~\cite{Guo2015,Alvarez2019,Huda2020,Maffei2018,Xie2019,Hetenyi2021}.
In a general one-dimensional metal with the Fermi wavenumer $k_F$, 
the charge susceptibility shows singularity at the wavenumber 
$Q=2k_F$ and the system has strong instability toward a charge order.
The Peierls phase transition and charge orders have been discussed in both organic materials and inorganic materials,
and there are a variety of charge orders with different modulations $Q$
~\cite{Kagoshima1981,Smaalen2004}.
In this context, the standard SSH and RM models correspond to the simplest charge order with $Q=\pi$. 
In spite of the extensive studies on the charge polarization, however, focus has been made mainly on this particular 
case and polarizations for other charge orders with general $Q$-vectors have not been well explored.
Since there are a lot of charge ordered one-dimensional materials~\cite{Kagoshima1981,Smaalen2004}, 
it is natural to consider generalization of the $Q=\pi$ order to $Q=2k_F\neq \pi$ orders.
Such generalization may be interesting also in the context of symmetry protected topological phases
beyond the non-interacting fermions.
For example, an $S=1/2$ trimer spin chain exhibits a 1/3 magnetization plateau implying 
a uniquely gapped ground state under a magnetic field,
which can be regarded as a generalization of more familiar spin dimer orders~\cite{Hida1994,Hu2014,Bera2022}. 
Modulated spin systems could have ``charge polarization" associated with 
the U(1) spin rotation symmetry similarly to 
charge ordered fermion systems~\cite{Oshikawa1997,Nakamura2002,Tasaki2018,Watanabe2021,Furuya2021}.

In this study, we investigate the charge polarization in a generalized RM model with a wavenumber 
$Q=2k_F$ at arbitrary filling per site $\nu$.
The model has neither the conventional bond-centered inversion symmetry nor the one site translation symmetry,
but has a combination of these symmetries.
The combined symmetry leads to quantization of the charge polarization, and 
it is independent of the parameters and determined solely by the filling $\nu$ in presence of the
symmetry.
Especially, the polarization can remain non-zero in the low density limit $\nu\to0$.
We also discuss excess charges accumulated near edges of the system under the open boundary condition.
Correspondingly, the polarization can be decomposed into a bulk contribution and an edge contribution.
It is found that these two contributions depend on parameters, while their sum remains constant 
under the symmetry.
Finally, we investigate effects of a phase shift in the modulation as a simple generalization of the model, 
where we introduce the intra unit cell and inter unit cell contributions of the
charge polarization and 
examine absence/existence of
gapless edge modes protected by the inversion symmetry.

\section{Quantized charge polarization under symmetry}
\begin{figure}[htb]
\includegraphics[width=8.0cm]{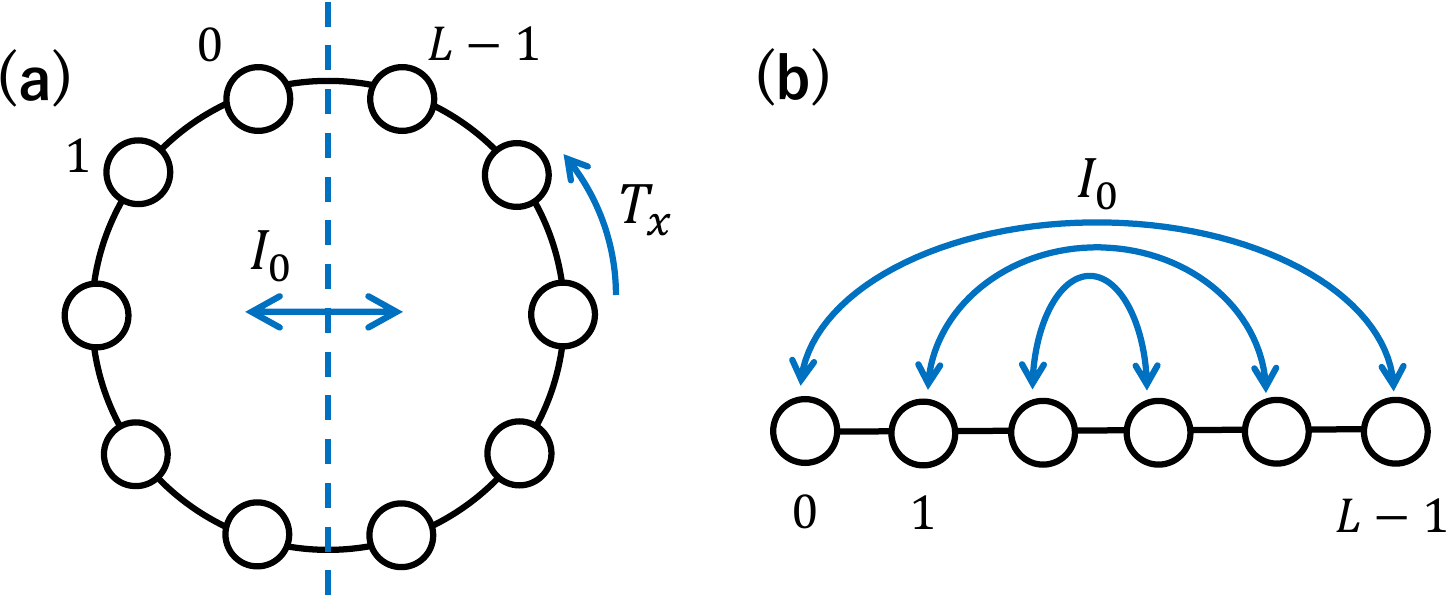}
\caption{Schematic pictures of the system for (a) the periodic boundary condition
and (b) the open boundary condition.
$I_0$ is the inversion about the bond $\braket{L-1,0}$ and $T_x$ is the one site translation.
The open boundary condition is realized by cutting the hopping integral on
the bond $\braket{L-1,0}$ in the periodic boundary condition.
}
\label{fig:system}
\end{figure}

\begin{figure}[htb]
\includegraphics[width=8.0cm]{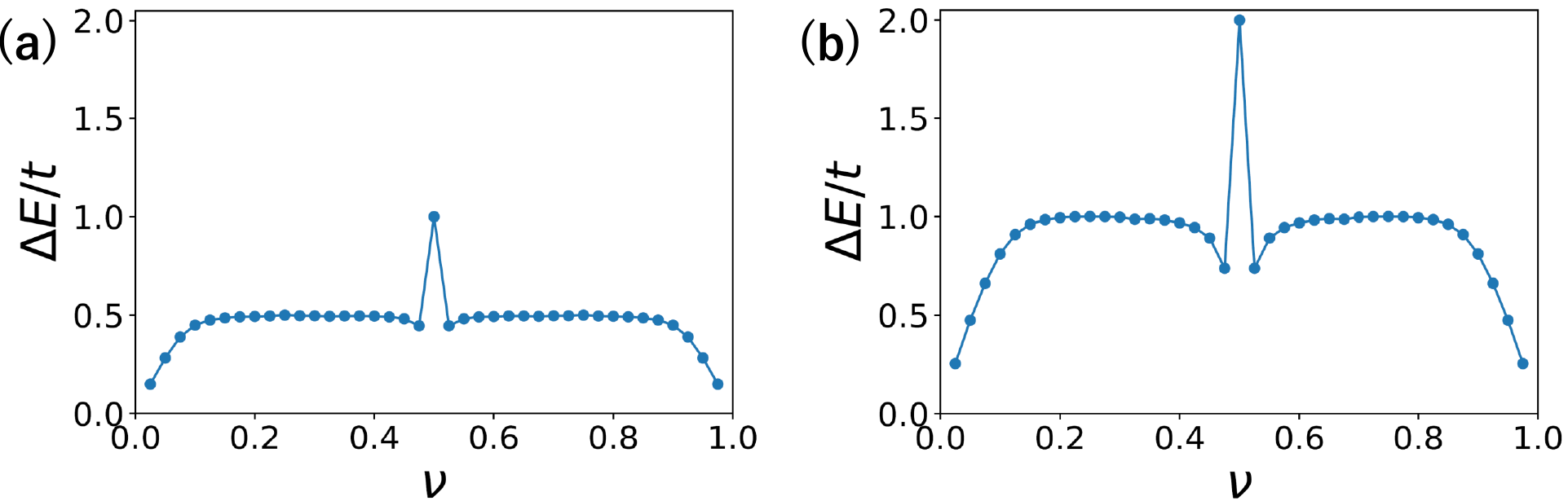}
\caption{Numerical results of the energy gap at (a) $\delta=0, \Delta=0.5t$
and (b) $\delta=0.5t,\Delta=0$ for the system under the periodic boundary condition.
}
\label{fig:gap}
\end{figure}

It is well known that a one dimensional fermion system has the Peierls instability
where the modulation period is determined by the Fermi wavevector or equivalently the particle density
~\cite{Kagoshima1981,Smaalen2004}.
In this study, we consider fermion systems with a modulated hopping integral and a modulated on-site potential
as a toy model for charge ordered insulators.
Such a system can be described by 
a generalized Rice-Mele model for spinless fermions~\cite{SSH1979,SS1981,RM1982,SSHbook,
Li2014,Yuan2017,Perez2019,Scollon2020,Guo2015,Alvarez2019,Maffei2018,Xie2019},
\begin{align}
H_0&=-\sum_{j}t_j(c_j^{\dagger}c_{j+1}+c_{j+1}^{\dagger}c_{j})+\sum_j\Delta_jc_j^{\dagger}c_{j}, 
\label{eq:H} \\
t_j&=t+\delta\cos(Qj), \quad \Delta_j=\Delta\cos(Qj),
\label{eq:cos}
\end{align}
where the sites are labeled as $j=0,1,2,\cdots,L-1$.
See Fig.~\ref{fig:system}.
The modulation wavenumber is given by $Q=2\pi \nu$, where $\nu=N/L=p/q$ is the particle filling
per site with $N$ being the total number of fermions and $p,q$ being coprime integers
($p$ fermions per $q$-site unit cell on average).
The system size is assumed to be an integer multiple of the unit cell for both the periodic and open
boundary conditions, $L=L_qq$, and is taken to be $L=600$
for numerical calculations in this study.
The modulation could be induced by interactions or arise from different atom species
in real materials,
but here $(\delta,\Delta)$ are model parameters of the system.
The modulation factor $\cos(Qj)$ reduces to the conventional factor $(-1)^j$ at the half-filling $\nu=1/2$.
One can introduce a phase shift such as $\cos(Q(j-j_0))$ with $j_0=1,2,\cdots,q-1$ 
for which the corresponding Hamiltonian
is related to the above $H_0$ by the $j_0$-site translation (see Sec.~\ref{sec:phase_shift}). 
It is straightforward to numerically calculate single-particle spectra of Eq.~\eqref{eq:H} and  
there exists an energy gap $\Delta E$
between the $N$-th and $(N+1)$-th single-particle energy eigenvalues under the periodic boundary condition
as shown in Fig.~\ref{fig:gap}.
One can add interactions such as $H_{\rm int}=\sum_{ij}V_{ij}n_in_j$ where $n_j=c_j^{\dagger}c_j$
and $V_{ij}$ keeps the symmetry of the system.
A large part of our discussions including 
symmetry-based arguments presented below is applicable also to interacting systems,
which will be important when one considers bosonic particles and spin systems. 
Note that our ground state is a uniquely gapped insulator with an integer filling per unit cell,
which is distinct from interaction-driven degenerate ground states~\cite{Aligia1999}.

Let us discuss symmetry of the Hamiltonian~\eqref{eq:H} and the resulting quantization of the charge polarization
under the periodic boundary condition.
Obviously, neither the one site translation symmetry $T_x: c_j\to c_{j+1}$
nor the bond-centered 
inversion symmetry $I_0: c_j\to c_{L-j-1}$ 
about the bond $\braket{L-1,0}$ alone is preserved for general $Q$ 
when $\delta$ and $\Delta$ are non-zero.
($\braket{j,j+1}$ denotes the bond between the sites $j$ and $j+1$.)
However, one can see that appropriate combinations of these symmetries are preserved.
Indeed, the Hamiltonian is invariant under the combined symmetry
when either $\delta$ or $\Delta$ is zero,
\begin{align}
I_k=(T_x)^kI_0=
\left\{
\begin{array}{ll}
T_xI_0 & (\delta=0,\Delta\neq0),\\
T_x^2I_0 & (\delta\neq0, \Delta=0).
\end{array}
\right.
\end{align}
Namely, the system has the $I_1$-symmetry (which is the site-centered inversion symmetry about the site
$j=0$) 
when $\delta=0$ and $I_2$-symmetry (which is the bond-centered inversion symmetry about 
the bond $\braket{0,1}$) when $\Delta=0$.
This can easily be shown as follows.
For the potential term, $I_0n_jI_0^{-1}=n_{L-j-1}$ and $T_x^kn_jT_x^{-k}=n_{j+k}$,
which leads to $T_x^kI_0\sum\cos(Qj)n_j(T_x^kI_0)^{-1}= \sum\cos(Q(-j+k-1))n_j$.
This implies $Qj=Q(j-k+1)$ in modulo $2\pi$ and hence $k=1$ (mod $q$).
For the hopping term, $I_0c_j^{\dagger}c_{j+1}I_0^{-1}=c_{L-j-1}^{\dagger}c_{L-j-2}$
and therefore $Qj=Q(j-k+2)$ in modulo $2\pi$, implying $k=2$ (mod $q$).
Note that the $I_k$-symmetry is just an inversion symmetry about an appropriate site or bond
translated by $T_x^k$ from the original inversion center $\braket{L-1,0}$,
but it turns out that clearly regarding $I_k$-symmetry as a combined symmetry is highly helpful.
The inversion center dependence will be essentially important for a system under the
open boundary condition.
A generalization for shifted modulations and inversion center dependence will be discussed in Sec.~\ref{sec:phase_shift}.

The combined symmetries can put strong constraints on the charge polarization
as in the standard SSH model at half filling $\nu=1/2$
where it is quantized to either $0$ or $1/2$ by the bond-centered inversion symmetry $I_0$.
Here we consider the charge polarization in the ground state $\ket{\Psi}$
based on the Resta formula which is equivalent to the Berry-Zak
formula in one dimension~\cite{Resta2007,Vanderbilt2018,Resta1998,Resta1999},
\begin{align}
P_x&=\frac{1}{2\pi}{\rm Im}\log\bra{\Psi}U_x\ket{\Psi},\\
U_x&=\exp\left( \frac{2\pi}{L}\sum_j j(n_j-\nu)\right).
\end{align}
The summand $(n_j-\nu)$ corresponds to the total electric charge at the site $j$
including the uniform ion contribution $-\nu$.
We note that the Resta formula itself is well-defined and can be used 
for both the periodic boundary condition and the open boundary condition.
The operator $U_x$ is often called Lieb-Schultz-Mattis twist operator
~\cite{Oshikawa1997,Nakamura2002,Bohm1949,LSM1961,Tasaki2018,Tasaki2021} 
and its expectation value in 
a uniquely gapped ground state is known to be $|\braket{U_x}|\simeq1$ in one dimension. 
This is because the energy of the variational state $\ket{\Phi}=U_x\ket{\Psi}$ converges to the ground state
energy in the thermodynamic limit
and therefore $\ket{\Phi}$ must be essentially proportional to the ground state
with a phase factor, $\ket{\Phi}\simeq e^{i2\pi P_x}\ket{\Psi}$, 
in a uniquely gapped system. See also Appendix~\ref{app:LSM}.
Thus the Resta formula for the polarization is well-defined in one dimensional uniquely gapped ground state,
although the expectation value is vanishing in the thermodynamic limit for general dimensions
and an appropriate modification is necessary to resolve such a problem
~\cite{WatanabeOshikawa2018,Tada2023}.

\begin{figure}[htb]
\includegraphics[width=8.0cm]{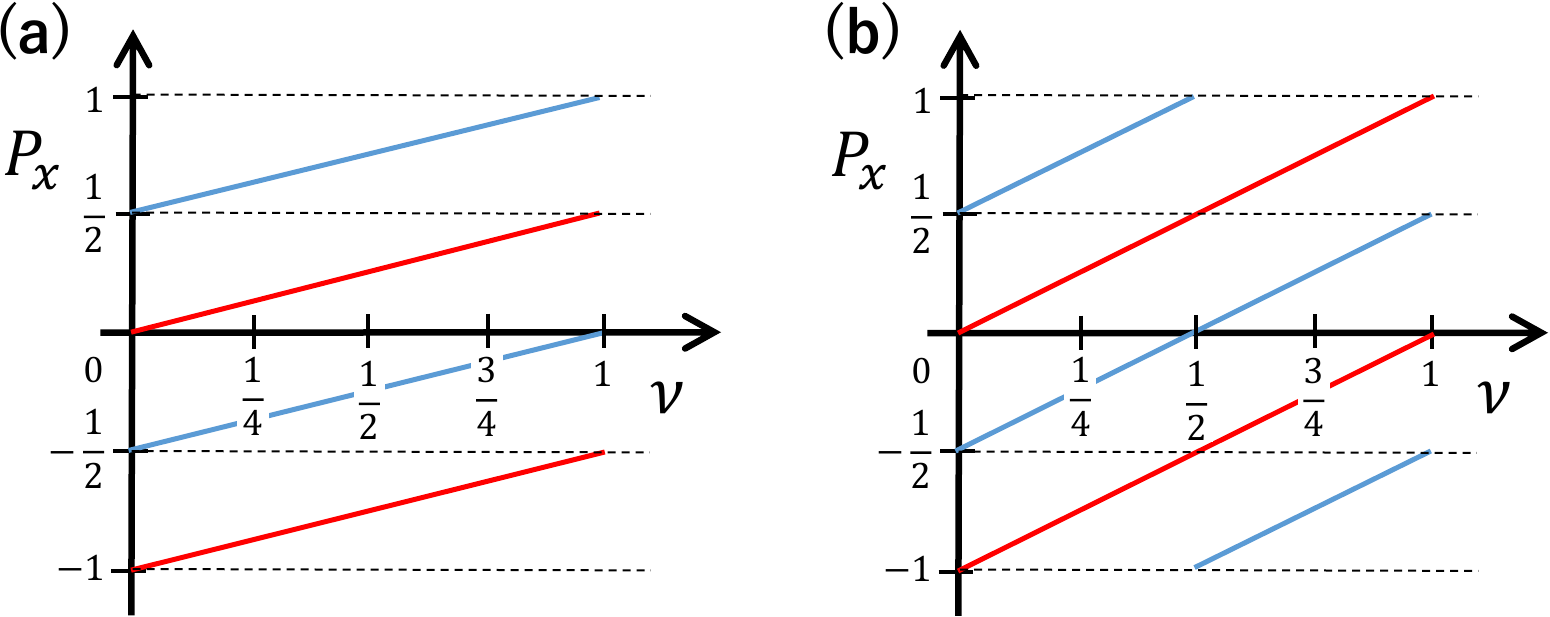}
\caption{The charge polarization quantized by (a) $I_1$-symmetry for $\delta=0, \Delta\neq0$
and (b) $I_2$-symmetry for $\delta\neq0,\Delta=0$.
The branches in the same color are equivalent, while those with different colors are inequivalent.
Only the region $-1\leq P_x\leq 1$ is shown for simplicity.
}
\label{fig:px}
\end{figure}

The Lieb-Schultz-Mattis operator behaves under the symmetries as
\begin{align}
I_0U_xI_0^{-1}&=U_x^{-1},\\
T_xU_xT_x^{-1}&=U_xe^{-i2\pi\nu}.
\end{align}
Note that these properties themselves are true for both the periodic boundary condition and the open boundary
condition, since they are just algebraic relations of the operators
and independent of a quantum state within the fixed filling Hilbert space.
The above equations imply that, for the uniquely gapped ground state $\ket{\Psi}$ which is an eigenstate of the
combined operator $I_k$ under the periodic boundary condition,
\begin{align}
e^{2\pi iP_x}=
\bra{\Psi}U_x\ket{\Psi}=e^{2\pi ik\nu}\bra{\Psi}U_x^{\dagger}\ket{\Psi}
=e^{2\pi i(k\nu-P_x)}.
\end{align}
Therefore, the charge polarization is quantized as 
\begin{align}
2P_x=k\nu \quad (\mbox{mod }1).
\label{eq:2P}
\end{align}
$P_x$ has two kinds of branches, namely, $P_x=k\nu/2 + n$
and $P_x=-1/2+k\nu/2+n$ with $n=0,\pm1,\pm2,\cdots$ as shown in Fig.~\ref{fig:px}.
For example, $P_x=+1/4$ and $P_x=-1/4$ at $\nu=1/2$ for $\delta=0$ are inequivalent.
Similarly, $P_x=0$ and $P_x=1/2$ at  $\nu=1/2$ for $\Delta=0$ correspond to distinct gapped states
of the standard SSH model.
We stress that Eq. \eqref{eq:2P} is valid in presence of an interaction under the $I_k$-symmetry
as long as the total Hamiltonian also preserves the symmetry and the many-body energy gap does not close.

Interestingly, the symmetry arguments claim that the polarization is $P_x\to 1/2$ (mod 1) in the limit of
zero-filling per site, $\nu\to0$, for the blue-colored branches in Fig.~\ref{fig:px}.
This may be counter-intuitive at first sight, but it does not contradict thermodynamics
because charge polarization is {\it not} a thermodynamic quantity.
The non-thermodynamic nature of the charge polarization is evident in a gedankenexperiment
where the polarization is easily changed by an additional tiny amount of charges put on a surface of a system.
Formally, there is no healthy thermodynamic limit under a uniform electric field as an intensive quantity
which is conjugate to the polarization, because the energy cost due to the corresponding scalar potential
is super-extensive similarly to the orbital angular momentum in superfluids~\cite{Tada2015,Tada2018}.
Nevertheless, it is widely considered that a fractional part of the polarization is an intrinsic or ``bulk"
quantity, and therefore one might naively expect that it vanishes in the low density limit.
This seemingly antagonistic problem can be resolved once one realizes that the charge polarization is not a standard
thermodynamic observable but is a kind of a center of mass of the particles.
The symmetry arguments simply mean that the center of mass in the present system can be $O(1)$ 
at $\nu=p/q\to0$, where there are $p$ fermions in each $q$-site unit cell on average.
Indeed, a single particle can have an $O(1)$ center of mass in the unit cell of large $q\gg 1$.
This can be confirmed by a straightforward calculation in the strong potential limit $|\Delta|\gg t$ at $\delta=0$,
where the Hamiltonian is reduced to $H=\sum\Delta\cos(Qj)n_j$.
The ground state of this Hamiltonian is simply given by a classical configuration of fermions which 
minimizes the potential $\Delta\cos(2\pi pj/q)$.
We suppose $p=1$ and $q\in 2{\mathbb Z}$ for simplicity, and then
one fermion is localized at every $j= 0$ (mod $q$) in the ground state when $\Delta<0$.
The polarization for this state is
\begin{align}
P_x&=
\frac{1}{2\pi}{\rm Im}\log \exp\left(i\frac{2\pi}{L}\sum_{k=0}^{L_q-1} kq -i\frac{2\pi}{L}\sum_{j=0}^{L-1} j\nu\right)
\nonumber\\
&= \frac{1}{2}\left(\frac{L}{q}-1\right)-\frac{1}{2q}(L-1)\nonumber\\
&=-\frac{1}{2}+\frac{\nu}{2} \quad (\mbox{mod } 1).
\end{align}
Note that the $L$-linear terms cancel out in the present charge neutral system,
and the final result is fully consistent with Eq.~\eqref{eq:2P}. 
Similarly, one fermion is located at every $j= q/2$ (mod $q$) when $\Delta>0$,
and thus
\begin{align}
P_x=\frac{\nu}{2} \quad (\mbox{mod } 1).
\end{align}
We stress that these results obtained in the limit $|\Delta|\to\infty$ are valid for general $\Delta$
as long as the energy gap does not close as $|\Delta|$ is decreased, because $P_x$ is quantized by the 
$I_k$-symmetry.

The above symmetry arguments are easily confirmed by numerical calculations.
We examplify numerical results of the Resta formula in Fig.~\ref{fig:px_num} (see Appendix~\ref{app:Ux} 
for technical details). 
They are fully quantized and characterized solely by the particle filling $\nu$ under the $I_k$-symmetry 
in exact agreement with the symmetry arguments.
As was shown above, the two inequivalent branches correspond to different signs of $\delta$ and $\Delta$.
The red-colored branches correspond to $\Delta>0$ in Fig.~\ref{fig:px_num} (a) and $\delta<0$ in (b),
while blue-colored branches to $\Delta<0$ in (a) and $\delta>0$ in (b).
\begin{figure}[htb]
\includegraphics[width=8.0cm]{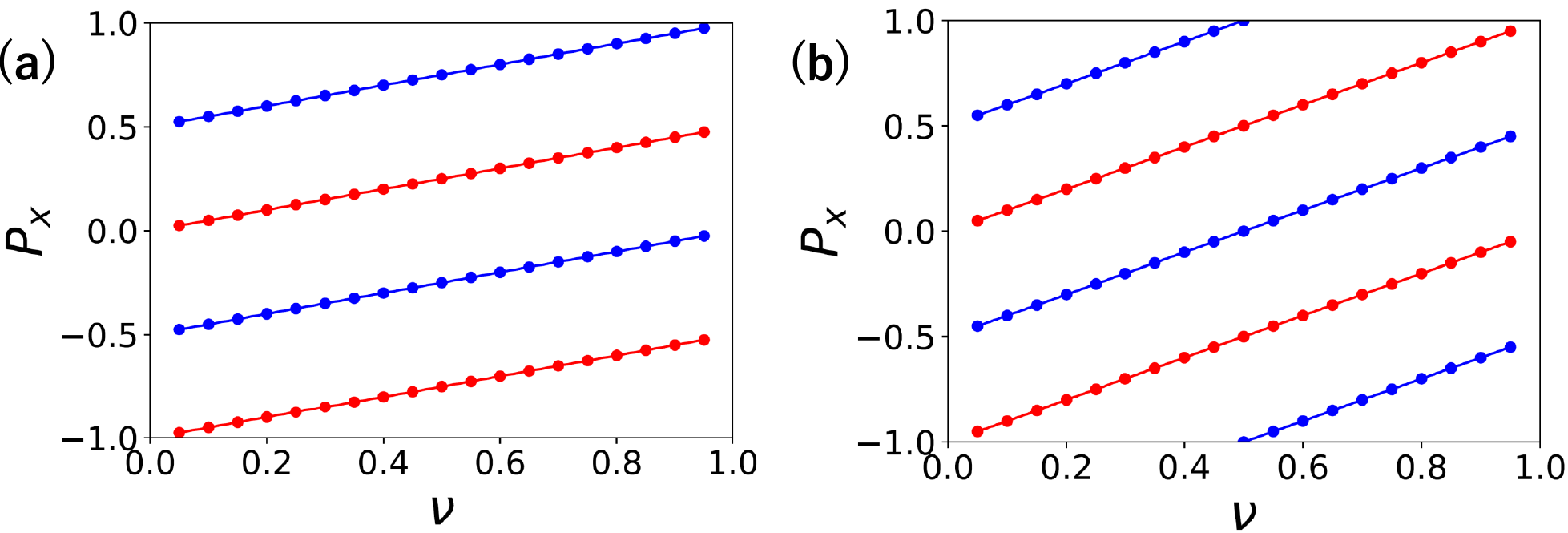}
\caption{Numerical results of the charge polarization at (a) $\delta=0, |\Delta|=0.5t$
and (b) $|\delta|=0.5t,\Delta=0$.
The branches in the same color are equivalent, while those with different colors are inequivalent.
Only the region $-1\leq P_x\leq 1$ is shown for simplicity.
}
\label{fig:px_num}
\end{figure}

\begin{figure}[htb]
\includegraphics[width=6.0cm]{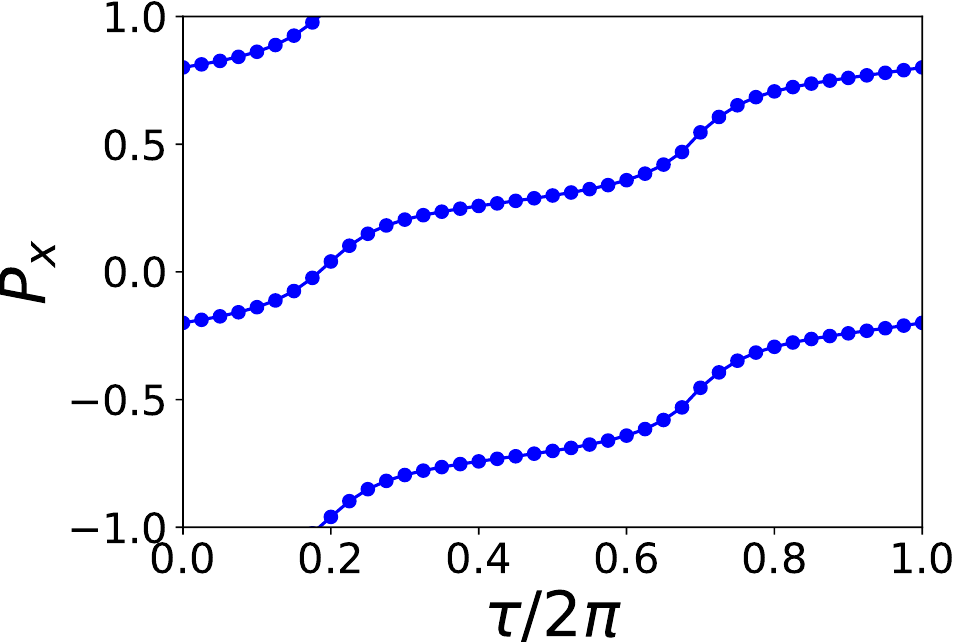}
\caption{Numerical demonstration of the charge pumping at $\delta_0=\Delta_0=0.5t$ calculated with
the Resta formula
for the system under the periodic boundary condition.
The filling is $\nu=3/10$.
}
\label{fig:pump}
\end{figure}

In general, both of $\delta$ and $\Delta$ can be non-zero simultaneously.
One can parameterize them as 
\begin{align}
\delta(\tau) &= \delta_0\cos\tau,\\
\Delta(\tau) &= \Delta_0\sin\tau.
\end{align}
We consider charge pumping during an adiabatic cycle $\tau=0\to2\pi$. 
An example of the calculated polarization in this process is shown
in Fig.~\ref{fig:pump}. 
The pumped charge $\Delta Q$ is quantized as
\begin{align}
\Delta Q=\int_0^{2\pi} \frac{dP_x}{d\tau} d\tau= 1,
\end{align}
as expected from the general theory of the Thouless pumping~\cite{Thouless1983,ThoulessPump2023}.
This is a straightforward generalization of the Thouless pumping in the standard RM model at $\nu=1/2$
to that at arbitrary filling.

\section{charge polarization under open boundary condition}
\begin{figure}[htb]
\includegraphics[width=8.0cm]{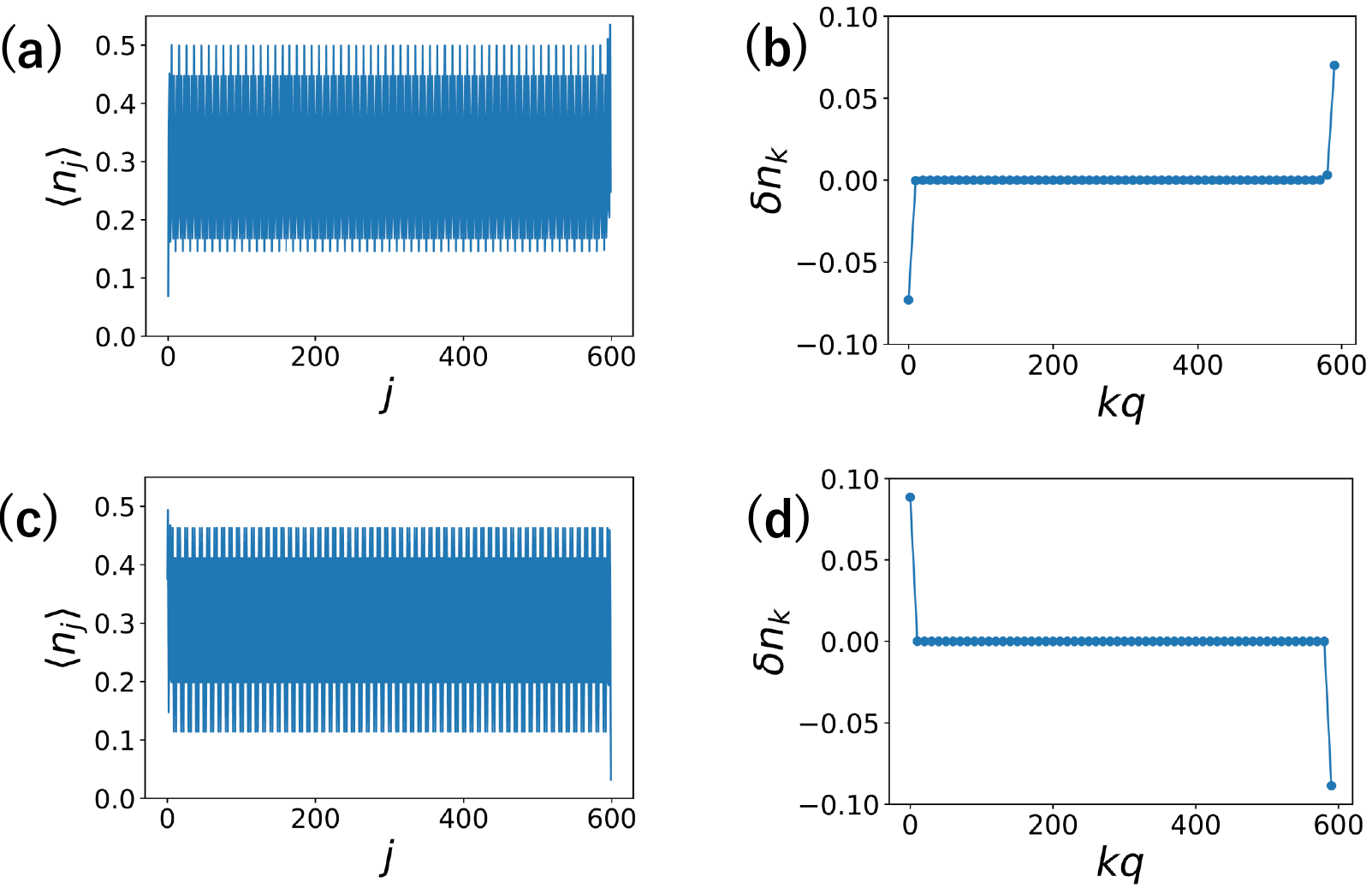}
\caption{The fermion particle density $\braket{n_j}$ and the excess charge $\delta n_k$
for (a), (b) $\delta=0, \Delta=0.5t$ and (c), (d) $\delta=0.5t, \Delta=0$
for the system under the open boundary condition.
The filling is $\nu=3/10$.
Note that $\braket{n_j}$ and $\delta n_k$ are not inversion symmetric since there is no inversion symmetry
in the system.
}
\label{fig:nj}
\end{figure}

In this section,
we consider the model Eq.~\eqref{eq:H} with the sites $j=0, 1,\cdots,L-1$
under the open boundary condition where there is no hopping 
between the right edge site $j=L-1$ and the left edge site $j=0$ (Fig.~\ref{fig:system}).
The inversion acts as $I_0: c_j\to c_{L-j-1}$, as in the periodic boundary condition.
In the standard SSH model ($\delta\neq0, \Delta=0$) at $\nu=1/2$, there can be gapless edge modes thanks to the 
chiral symmetry, but existence of such single-particle edge modes are not generally guaranteed 
in a system only with point group symmetry~\cite{Song2017,Huang2017}.
Similarly, there are no edge modes in the present system at almost all filling
except for some special cases which will be discussed in Sec.~\ref{sec:phase_shift}.
Nevertheless, 
one may naively expect that there are extra charges accumulated 
near edges of the system for a given filling. 
Here we introduce excess charges over one unit cell (length $q$) compared to the uniform background
charges,
\begin{align}
\delta n_k=\sum_{j=kq}^{(k+1)q-1}(\braket{n_j}-\nu).
\end{align}
The excess charge $\delta n_k$ corresponds to 
the total charge of electrons and ions contained in the $k$-th unit cell.
Figure~\ref{fig:nj} shows examples of the fermion density profiles $\braket{n_j}$ and 
the corresponding excess charges $\delta n_k$. 
Clearly, the excess charges vanish in the bulk region of the system, although the fermion density itself oscillates 
with the period $q$.
This is a general property for a gapped periodic system which has unit cell translation symmetry.
There are excess edge charges localized around the ends of the system and they are
quantified as
\begin{align}
Q_{\rm left}&=\sum_{k=0}^{k_e}\delta n_k, \quad Q_{\rm right}=\sum_{k=L_q-1-k_e}^{L_q-1}\delta n_k.
\end{align}
$k_e$ specifies an edge region of length $O(1)$ 
and is supposed to be sufficiently larger than the correlation length of the system.
In the numerical calculations, however,
it is chosen as $k_e=[L_q/4]$ for numerical stability when the energy gap
is small or equivalently the correlation length is large.
We can numerically confirm $Q_{\rm left}+Q_{\rm right}=0$ holds because of the charge neutrality in the entire system,
although $\delta n_k$ itself is not inversion symmetric in general.
(Typically, $|Q_{\rm left}+Q_{\rm right}|=O(10^{-8\sim -7})$ for the system size $L=600$.
The equation $Q_{\rm left}+Q_{\rm right}=0$ is satisfied in the thermodynamic limit, 
$\lim_{k_e\to\infty}\lim_{L\to\infty}(Q_{\rm left}+Q_{\rm right})=0$,
because $Q_k\simeq 0$ for the $k$-th unit cell in the bulk region in a gapped system.)
We emphasize that
the present excess edge charges are different from the (smeared or weighted)
edge charges widely discussed in the literatures
~\cite{Resta2007,Vanderbilt2018,Vanderbilt1993,Watanabe2021,Furuya2021}.
The smeared edge charge is given by a convolution of the charge density with a broadening function and it
exactly equals the polarization.
On the other hand,
our excess edge charge only partly contributes to the polarization as will be discussed in the following.
Our excess charge in the present lattice model would correspond to the extra charge
in Bloch electron systems discussed in the previous study~\cite{Rhim2017}.
It may be regarded as a many-body variant of the previously introduced extra edge charge
based on the trivial decomposition of the Resta formula.

Let us now discuss the polarization under the open boundary condition.
There will be contributions to the polarization from the edge region with width $O(1)$, when the
excess edge charges are non-zero.
At the same time, microscopic charge configuration within a unit cell will also contribute to the polarization
and such a contribution comes uniformly from a bulk region of the system.
Correspondingly, we introduce a bulk contribution and an edge contribution under the open boundary condition
by decomposing the polarization operator.
The polarization operator is given by
\begin{align}
\hat{P}_x=\frac{1}{L}\sum_{j=0}^{L-1} j(n_j-\nu)
\end{align}
and it is well-defined under the open boundary condition.
For a state with the above mentioned charge configuration, the expectaion value of this operator is simplified as
\begin{align}
\braket{\hat{P}_x}&=\frac{1}{L}\sum_{k=0}^{L_q-1}\sum_{j=0}^{q-1}(kq+j)(\braket{n_{kq+j}}-\nu) \nonumber\\
&\simeq \frac{1}{q}\sum_{j=0}^{q-1}j(\braket{n_{kq+j}}-\nu)\big|_{k\gg1 } \nonumber\\
&\qquad +\frac{1}{L}\sum_{k\simeq0, L_q}kq\sum_{j=0}^{q-1}(\braket{n_{kq+j}}-\nu).
\label{eq:decompose}
\end{align}
Therefore, the expectation value of the polarization operator is simply evaluated as
\begin{align}
\braket{\hat{P}_x}&\simeq P_{\rm bulk}+P_{\rm edge},\\
P_{\rm bulk}&=\frac{1}{q}\sum_{j=j_b}^{j_b+q-1}j(\braket{n_j}-\nu),\\
P_{\rm edge}&=(Q_{\rm right}-Q_{\rm left})/2.
\end{align}
$j_b=nq (n\gg 1)$ defines a bulk region and is chosen as $n=[L_q/4]$ in the numerical calculations.
This is an almost trivial decomposition and is valid for general systems with gapped bulk excitations, 
for which effects of a boundary condition
are exponentially localized only around an edge.
In such a system, the edge contribution can be evaluated under the periodic boundary condition
as $P_{\rm edge}=P_x-P_{\rm bulk}$ in modulo 1, 
where $P_x$ and $P_{\rm bulk}$ are calculated under the periodic boundary condition 
(see also Sec.~\ref{sec:phase_shift}). 
Therefore, the excess edge charge is regarded as an intrinsic quantity of the system,
which holds true not only in non-interacting models but also in interacting models.

\begin{figure}[htb]
\includegraphics[width=8.0cm]{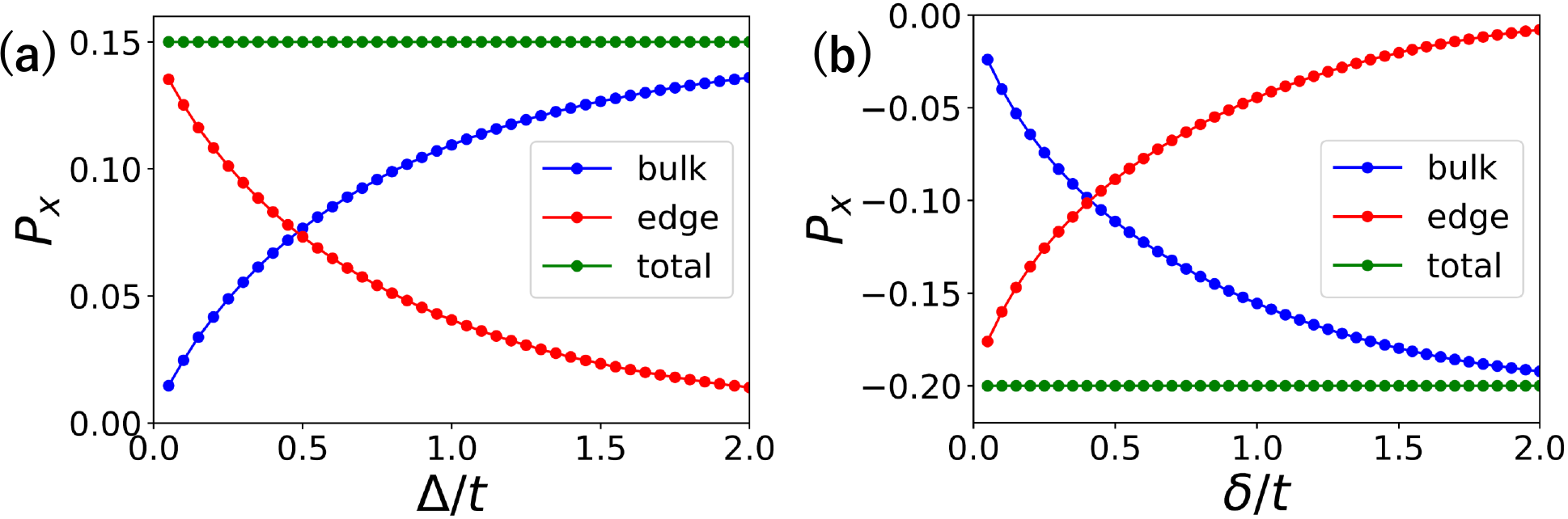}
\caption{The bulk (blue) and edge (red) contributions of the charge polarization
for (a) $\delta=0$ and (b)$\Delta=0$
for the system under the open boundary condition.
The filling is $\nu=3/10$.
The total polarization (green) is the numerical sum of these two contributions.
}
\label{fig:Pbet}
\end{figure}

In Fig.~\ref{fig:Pbet},
we examplify numerical results of the bulk and edge contributions when one of $\delta$ and $\Delta$ is zero.
Each of the contributions depends on the parameters, but their sum remains constant
and is quantized under the $I_k$-symmetry.
We numerically see that 
\begin{align}
\braket{\hat{P}_x}\simeq P_{\rm bulk}+P_{\rm edge}\simeq P_x \quad (\mbox{mod }1),
\label{eq:PP}
\end{align}
where $P_x$ is the polarization evaluated with use of the Resta formula under the open
boundary condition. (As mentioned in the previous section, the Resta formula is 
well-defined both for the periodic boundary condition and the open boundary condition.)
Note that $P_x$ under the open boundary condition numerically coincides with 
the one under the periodic boundary condition,
which would be a general property as discussed in Appendix~\ref{app:BC}.
We confirm that
typical difference between $\braket{\hat{P}_x}, (P_{\rm bulk}+P_{\rm edge})$, and 
$P_x$ is small ($O(10^{-4\sim-3})$ when $L=600$), and it decreases as the system size increases
in the numerical calculations.
Although the total polarization is a quantized constant under the symmetry,
each of $P_{\rm bulk}$ and $P_{\rm edge}$ changes continuously with the parameters.
The bulk contribution decreases as the modulation strength is reduced
and approaches zero, $P_{\rm bulk}\to0$, when $\delta\to0$ or $\Delta\to0$, 
because the system becomes a simple metal with a uniform bulk charge density
in that limit.
At the same time, the edge contribution increases and approaches the quantized value,
which means that the excess edge charges $Q_{\rm left/right}$
become quantized in the weak modulation limit under the $I_k$-symmetry.
Similar quantization was discussed previously, but the fundamental difference is that 
the edge charges in the previous studies are defined so that they are equal to the total polarization for 
any parameters~\cite{Resta2007,Vanderbilt2018,Vanderbilt1993,Watanabe2021,Furuya2021}.
In contrast to the weak modulation limit, 
the total polarization is dominated by the bulk contribution while the edge contribution is decreased 
when the modulation is large, because the system is deep in an insulating regime.
In the strong modulation limit ($|\delta|\to\infty$ or $|\Delta|\to\infty$),
the edge contribution vanishes, $P_{\rm edge}\to0$.
These behaviors are generally seen except for special cases which will be discussed in the next section.

\section{Genralization
and bulk-boundary correspondence}
\label{sec:phase_shift}
The discussions on the model Eq.~\eqref{eq:H} in the previous sections can be generalized to some extent.
In this section, we discuss a simple generalization of the model when either $\delta$ or $\Delta$ is zero
and examine existence/absence of gapless edge modes protected by the inversion symmetry.
We introduce a phase shift in the cosine term and 
focus on a ``commensurate" phase shift $\cos(Q(j-j_0))$ where $j_0=1,2,\cdots, q-1$
for the filling $\nu=p/q$.
This modulation is also reduced to the standard staggered factor $(-1)^{j-j_0}$ at the half filling $\nu=1/2$.
Under the periodic boundary condition,
the Hamiltonian $H_{j_0}$ with $\cos(Q(j-j_0))$ is related to $H_0$ with $\cos(Qj)$ by the $j_0$-site translation,
\begin{align}
H_{j_0}=T_x^{j_0}H_{0}T_x^{-j_0}.
\label{eq:Hj0}
\end{align}
This means that the Hamiltonian $H_{j_0}$ has the $I_{k+2j_0}$-symmetry
when $H_0$ has the $I_k$-symmetry.
Besides, the energy spectra for these Hamiltonians are unitary equivalent and especially
the ground states are uniquely gapped under the periodic boundary condition. 
Thanks to the $I_{k+2j_0}$-symmetry, the charge polarization is quantized as
\begin{align}
2P_x=(k+2j_0)\nu \quad \mbox{(mod 1)}.
\end{align}
This is a natural result for the potential $\Delta\cos(Q(j-j_0))$, 
because the phase shift will change the positions of the potential minima by $j_0$ sites
and correspondingly the center of mass of the particles will move by $j_0\nu$ to minimize the ground state energy.
Note that the Hamiltonian has the $q$-site (one unit cell) translation symmetry and 
the center of mass comes to
the position which  is equivalent to the original position under the $q$-site translation.
A similar thing will take place for the system with the phase shifted hopping modulation, 
although it is not easy to provide an intuitive picture.

\begin{figure}[htb]
\includegraphics[width=8.0cm]{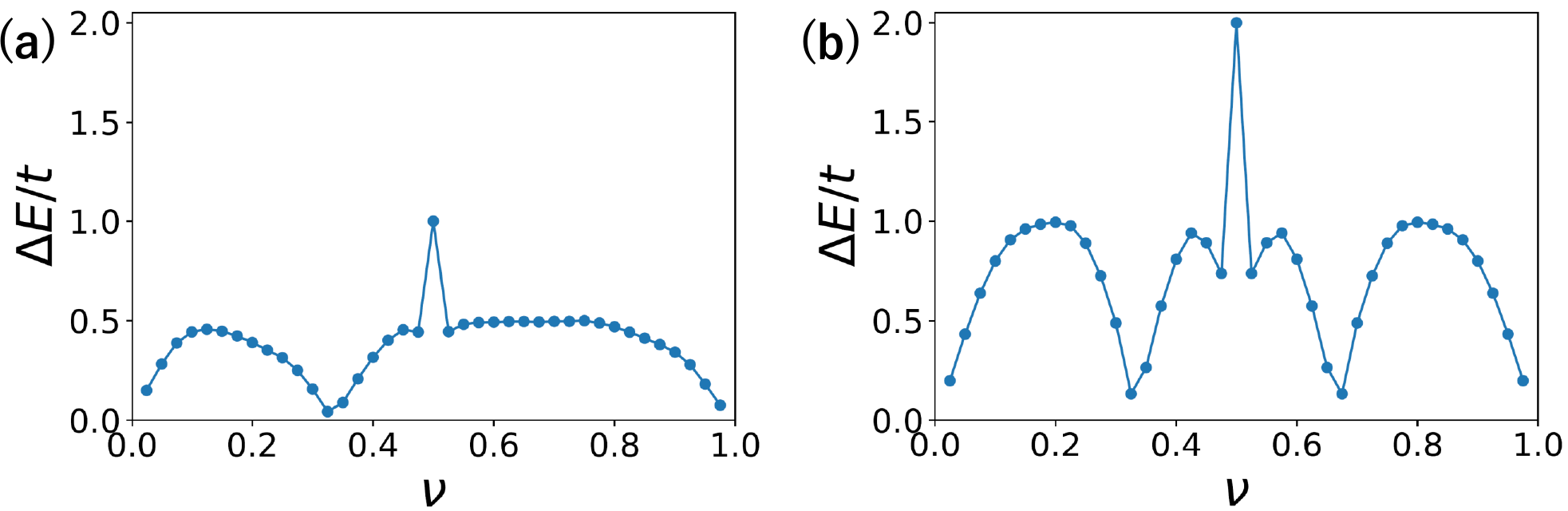}
\caption{Numerical results of the energy gap at (a) $j_0=1,\delta=0, \Delta=0.5t$
and (b) $j_0=2,\delta=0.5t,\Delta=0$ for the system under the open boundary condition.
The gap closing takes place at (a) $\nu^*=1/3$ and (b) $\nu^*=1/4,3/4$.
}
\label{fig:gapless}
\end{figure}

\begin{figure}[htb]
\includegraphics[width=8.0cm]{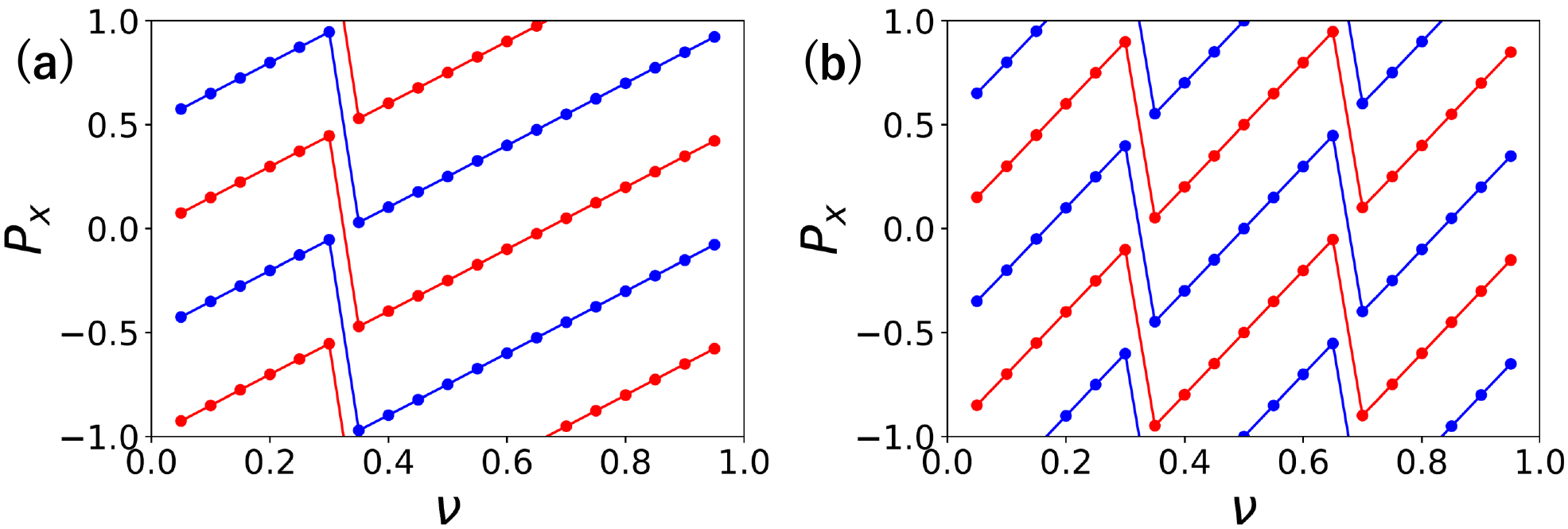}
\caption{Numerical results of the charge polarization at (a) $j_0=1,\delta=0, |\Delta|=0.5t$
and (b) $j_0=2,|\delta|=0.5t,\Delta=0$ for the system under the open boundary condition.
The solid lines are drawn with an emphasis on the gap closing for the eyes.
}
\label{fig:px_shift}
\end{figure}

\begin{figure}[htb]
\includegraphics[width=8.0cm]{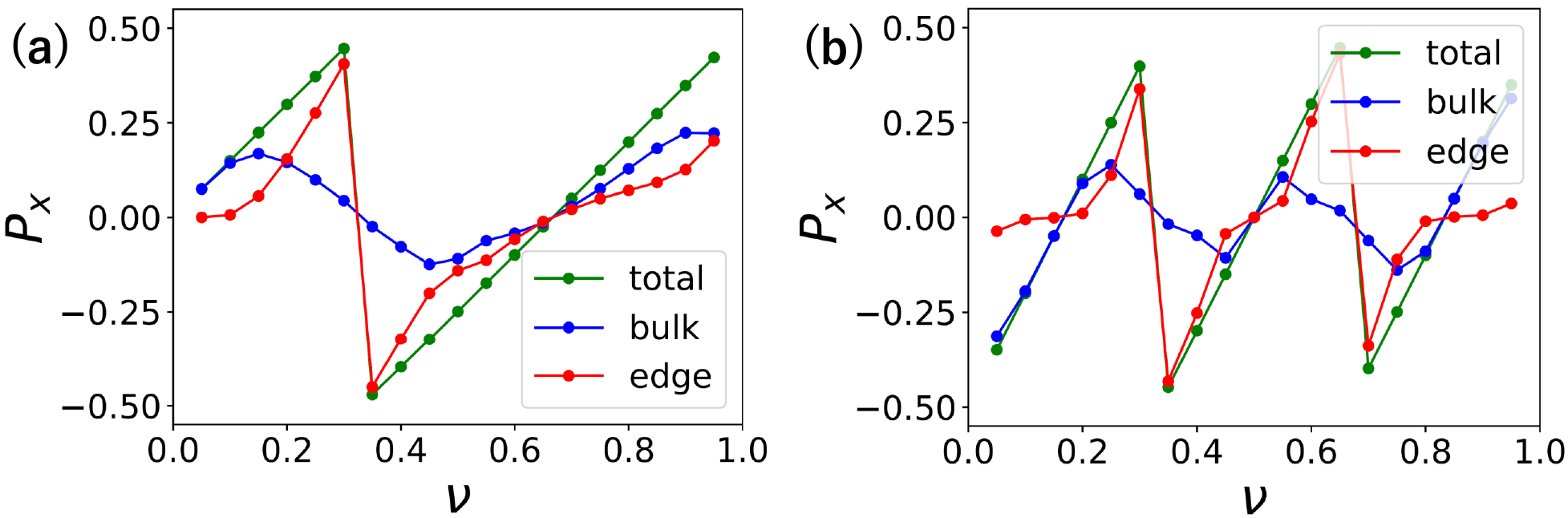}
\caption{The bulk (blue) and edge (red) contributions of the charge polarization
for (a) $j_0=1,\delta=0,\Delta=0.5t$ and (b)$j_0=2,\delta=0.5t,\Delta=0$
for the system under the open boundary condition.
The total polarization (green) is the numerical sum of these two contributions.
The bulk contribution vanishes and $P_x=P_x^{\rm edge}=\pm 1/2$ holds (a) at $\nu^*=1/3$
and (b) $\nu^*=1/4, 3/4$.
}
\label{fig:Pbet_shift}
\end{figure}

Under the open boundary condition where there is no hopping integral on the bond $\braket{L-1,0}$, 
Eq.~\eqref{eq:Hj0} does not hold and
the Hamiltonians $H_{j_0}$ and $H_0$ are not unitary equivalent.
In this case, we numerically find that the energy spectrum
of $H_{j_0}$ becomes
gapless at particular filling $\nu=\nu^*$ implying existence of gapless edge modes as shown in Fig.~\ref{fig:gapless}.
Alternatively, 
existence or absence of an gapless edge mode depends on the position of cutting a bond of the system
(setting the hopping integral to zero) to realize an open boundary condition.
The gap closing takes place when the filling satisfies the condition
\begin{align}
P_x = \frac{1}{2} \quad \mbox{(mod 1)},
\label{eq:P1/2}
\end{align}
or equivalently $(k/2+j_0)\nu^*=0, 1/2$ (mod 1) depending on the parameters as shown in Fig.~\ref{fig:px_shift}.
This includes the well-known gapless edge mode in the topologically non-trivial state of the SSH model at $\nu=1/2$
($\delta<0,\Delta=0,j_0=0$ in our model).
It is easy to see that,
under this condition, the Hamiltonian $H_{j_0}$ has the conventional inversion symmetry $I_0$ 
about the bond $\braket{L-1,0}$, $I_0H_{j_0}I_0^{-1}=H_{j_0}$.
In presence of this inversion symmetry, the bulk contribution of the polarization is zero, 
$P_x^{\rm bulk}=(1/q)\sum_{j=L-j_b-q}^{L-j_b-1}(L-1-j)(\braket{n_j}-\nu)\simeq -P_x^{\rm bulk}\simeq0$,  
because the summation of $(\braket{n_j}-\nu)$ over one unit cell in the bulk should be exponentially small.
Therefore, Eq.~\eqref{eq:P1/2} simply means $P_x=P_x^{\rm edge}=\pm1/2$, which can be confirmed numerically
as seen in Fig.~\ref{fig:Pbet_shift}. 
Strictly speaking, the inversion symmetric ground states under the open boundary condition 
are nearly degenerate
and $Q_{\rm left}=Q_{\rm right}=0$ for each state due to the inversion symmetry and charge neutrality.
$P_x^{\rm edge}=\pm1/2$ should be understood as either the limit $\nu\nearrow \nu^*$ or $\nu\searrow\nu^*$.

The above discussions are consistent with 
the general argument for Bloch band systems in the previous study~\cite{Rhim2017}.
Here, we provide a brief explanation of the bulk-boundary correspondence based on the Resta formula for 
the many-body (interacting) ground state.
Similary to Eq.~\eqref{eq:decompose}, 
we decompose the Lieb-Schultz-Mattis operator as $U_x=U_{\rm intra}U_{\rm inter}$, where
\begin{align}
U_{\rm intra}&= e^{i2\pi \hat{P}_x^{\rm intra}},\quad
\hat{P}_x^{\rm intra}= \frac{1}{L_q}\sum_{k=0}^{L_q-1}\frac{1}{q}\sum_{l=0}^{q-1}l(n_{kq+l}-\nu),\\
U_{\rm inter}&= e^{i2\pi \hat{P}_x^{\rm inter}},\quad
\hat{P}_x^{\rm inter}= \frac{1}{L_q}\sum_{k=0}^{L_q-1}k\sum_{l=0}^{q-1}(n_{kq+l}-\nu).
\end{align}
We call them the intra unit cell contribution and inter unit cell contribution, respectively.
Note that the bond $\braket{L-1,0}$ is an inter unit cell bond under the periodic boundary condition and 
the sites $j=0$ and $j=L-1$ are the edges under the open boundary condition in the present study.
The intra (inter) unit cell contribution under the periodic boundary condition 
corresponds to the bulk (edge) contribution $P_{\rm bulk} (P_{\rm edge})$ under the
open boundary condition. 
Indeed, it turns out that
\begin{align}
P_{\rm intra}=P_{\rm bulk}, \quad P_{\rm inter}=P_{\rm edge},
\end{align}
where
$P_a={\rm arg}\bra{\Psi}U_a\ket{\Psi}/2\pi (a=\mbox{intra, inter})$.
This is because the intra unit cell contribution has no fluctuations in a uniquely gapped ground state 
(see the discussion in the next paragraph and Appendix~\ref{app:Ux}) and thus $P_{\rm intra}=P_{\rm bulk}$ just follows.
Besides, the total polarization is independent of the boundary conditions (in modulo 1) 
as mentioned in the previous section
and Appendix~\ref{app:BC}, which means that $P_{\rm inter}=P_x-P_{\rm intra}
=\braket{\hat{P}_x}-P_{\rm bulk}=P_{\rm edge}$.

The variational energy is $\bra{\Phi_a}H_{j_0}\ket{\Phi_a}=O(1/L)$ for each of the states
$\ket{\Phi_a}=U_a\ket{\Psi}$ simiarly to the original Lieb-Schultz-Mattis operator
~\cite{Bohm1949,LSM1961}.
Correspondingly, deviations of the variational states from the uniquely gapped ground state 
under the periodic boundary condition are
$||\ket{\delta\Psi_a}||^2=O(1/L)$
when written as $\ket{\Phi_a}=e^{i2\pi P_a}\ket{\Psi}+\ket{\delta \Psi_a}$ (see Appendix~\ref{app:LSM}). 
The ground state is almost an eigenstate of $U_a$ in this sense, and thus the polarization is given by
$\braket{U_x}=\braket{U_{\rm intra}}\braket{U_{\rm inter}}+O(1/L)$, where 
$\braket{\cdots}=\bra{\Psi}\cdots\ket{\Psi}$.
An important point is that $\hat{P}_x^{\rm intra}$ is a normalized extensive operator whose norm is $O(1)$,
although $\hat{P}_x^{\rm inter}$ is super-extensive with the norm $O(L)$.
As a result, there are no fluctuations of the intra unit cell contribution and 
$\braket{U_{\rm intra}}=e^{i2\pi P_{\rm intra}}+O(1/L)\simeq 1$ holds,
because $(\partial/\partial\theta)\braket{U_{\rm intra}(\theta)}=
i\braket{\hat{P}_x^{\rm intra}}\braket{U_{\rm intra}(\theta)}+i\bra{\Psi}\hat{P}_x^{\rm intra}\ket{\delta\Psi_{\rm intra}} 
=0+O(1/\sqrt{L})$ 
for the inversion symmetric ground state
with $\braket{\hat{P}_x^{\rm intra}}=0$,
where $U_{\rm intra}(\theta)=e^{i\theta \hat{P}_x^{\rm intra}}$ (see also Appendix~\ref{app:Ux}).
On the other hand, there can be non-negligible fluctuations for the inter unit cell contribution,
since $\hat{P}_x^{\rm inter}$ is super-extensive.
Therefore, the charge polarization is dominated by the inter unit cell contribution for the
inversion symmetric uniquely gapped ground state in the thermodynamic limit,
\begin{align}
\braket{U_x}=\braket{U_{\rm inter}}, \quad P_x=P_{\rm inter}.
\end{align}

The inter unit cell Lieb-Schultz-Mattis operator has been used in the previous study,
where it was shown that a non-trivial expectation value $\braket{U_{\rm inter}}=-1$ in the
inversion symmetric ground state (for an infinite system) 
implies existence of a gapless edge state in presence of an edge (for a semi-infinite system)
~\cite{Tasaki2018,Tasaki2021}.
Roughly speaking,
a local variant of the inter unit cell Lieb-Schultz-Mattis operator $U_{\rm loc}(x)$ 
was introduced to characterize 
nature of a finite large region around a point $x\in{\mathbb R}$, where 
its expectaion value is $\pm1$ under the inversion symmetry.
The condition $\braket{U_{\rm loc}(x)}=-1$ for $x$ in the bulk of the system
together with the trivial condition $\braket{U_{\rm loc}(x)}=+1$
for $x$ outside of the system implies that there must be gap closing around the edge of the system.
This index theorem was rigorously proved as bulk-boundary correspondence
for an infinite system without an edge and a semi-infinite system
with an edge~\cite{Tasaki2018,Tasaki2021},
but we assume that it is applicable also to a finite large system with the periodic boundary condition and 
the same system with the open boundary condition.
Thus, we can summarize the above argument on the bulk-boundary correspondence for a (interacting) 
uniquely gapped state as follows.
\begin{claim}
In an inversion symmetric insulator,
there exists a gapless edge state under the open boundary condition if the charge polarization 
is $P_x=1/2$ (mod 1) under the periodic boundary condition.
\end{claim}

Finally, we remark that
the existence of a gapless (single-particle) edge mode or 
more generally degenerate ground states under the open boundary 
condition could also be understood as the filling anomaly 
~\cite{Benalcazar2019,Tada2023}.
We discuss the ground state wavefunctions when $\delta=0, \Delta\gg t$ in view of the filling anomaly
to have an intuitive understanding.
Let us consider a simple case with $j_0=1$ for which $P_x=1/2$ is acheived when $\nu=1/3 (p=1,q=3)$.
There are degenerate potential minima at $j=0$ and $j=2$ (mod 3) in a unit cell and the middle site 
$j=1$ (mod 3) corresponds to the potential maximum. 
Here, we define a unit cell as the sites $\{3k, 3k+1,3k+2\}, k=0,1,2\cdots,L_3-1$ for which the bonds
$\braket{3k-1,3k}$ including
$\braket{L-1,0}$ connect neighboring unit cells, where $L_3=L/3\in{\mathbb Z}$.
The ground state under the periodic boundary condition is 
\begin{align}
\ket{\Psi}&\simeq \frac{1}{\sqrt{2}}(c_{L-1}^{\dagger}+c_{0}^{\dagger})
\prod_{k=0}^{L_3-2}\frac{1}{\sqrt{2}}(c_{3k+2}^{\dagger}+c_{3k+3}^{\dagger})\ket{0}.
\label{eq:Psi3}
\end{align}
A dimer $(c_j^{\dagger}+c_{j+1}^{\dagger})\ket{0}$ is formed by the hopping $t>0$
between neighboring unit cells, which
is analogous to the standard SSH model with $P_x=1/2$ where there are dimers on inter unit cell bonds.
(See Appendix~\ref{app:psi3} for direct calculations of the polarization in the ground state \eqref{eq:Psi3}.)
The local excitated state on the dimer bond $\braket{j,j+1}$ 
is $(c_j^{\dagger}-c_{j+1}^{\dagger})\ket{0}$ with an energy gap $2t$ which has a different inversion eigenvalue 
from that of the ground state.
Under the open boundary condition without a hopping integral on the
bond $\braket{L-1,0}$, the energy gap between the two states $(c_{L-1}^{\dagger}+c_{0}^{\dagger})\ket{0}$
and $(c_{L-1}^{\dagger}-c_{0}^{\dagger})\ket{0}$ vanishes and the state 
\begin{align}
\ket{\Psi'}&\simeq \frac{1}{\sqrt{2}}(c_{L-1}^{\dagger}-c_{0}^{\dagger})
\prod_{k=0}^{L_3-2}\frac{1}{\sqrt{2}}(c_{3k+2}^{\dagger}+c_{3k+3}^{\dagger})\ket{0}
\end{align}
becomes another ground state.
(To be precise,
there is an exponentially small energy gap between these two ground states in a finite size system, but
they are exactly degenerate in the thermodynamic limit.)
We can also consider linear combinations of these states which explicitly break the inversion symmetry,
$\ket{\Psi_{\rm left}}=(\ket{\Psi}-\ket{\Psi'})/\sqrt{2}$ and $\ket{\Psi_{\rm right}}=(\ket{\Psi}+\ket{\Psi'})/\sqrt{2}$.
The state $\ket{\Psi_{\rm left}} (\ket{\Psi_{\rm right}})$ has a localized fermion at the left (right) edge $j=0 (j=L-1)$,
corresponding to the excess edge charges $Q_{\rm left/right}=\pm 1/2$.
We stress that
it is impossible to realize a uniquely gapped ground state under the open boundary condition
with keeping the inversion symmetry and the fermion particle number.
Essentially the same argument applies to other filling as well.
This is a filling anomaly~\cite{Benalcazar2019} and 
it can be discussed more rigorously with use of a polarization index~\cite{Tada2023}.

\section{discussion and summary}
We have studied the charge polarization in the generalized RM model in one dimension.
Motivated by the various charge ordered materials,
we introduced modulation wavenumber $Q=2\pi\nu$ with filling $\nu$ per site
corresponding to a charge order driven by the Peierls instability.
The model has the combination of the bond-centered inversion and one site translation
symmetries when either the hopping modulation or the potential modulation is zero.
This symmetry leads to the quantization of the charge polarization.
Especially, the charge polarization can be 1/2 (mod 1) at the low density limit $\nu\to0$.
These results are confirmed by the numerical calculations of the non-interacting model,
and Thouless pumping at arbitrary filling was also demonstrated.
Under the open boundary condition, there exist excess charges accumulated near the edges of the system.
Correspondingly, the charge polarization can be decomposed into the bulk and edge contributions.
The sum of these two contributions is a quantized constant under the combined symmetry,
while each of them depends on the modulation strength.
The polarization is dominated by the edge (bulk) contribution in the weak (strong) modulation limit.
We also discussed the effects of a phase shift and examined the absence/existence of gapless edge modes
or more generally degenerate ground states.
Especially, we introduced the intra and inter unit cell contributions of the charge polarization
 in the Resta formula
and clarified the role of the inversion symmetry.

A large part of our discussions including
the symmetry argument is applicable to general uniquely gapped systems beyond the non-interacting fermions
such as interacting fermions, bosons and spin systems.
For example, 
it would be interesting to study spin systems with large unit cells such as 
spin Peierls systems and trimer spin chains
in view of the $I_k$-symmetry protected topological state.
Our work could provide an insight for a further development in the understanding of the charge polarization.

\begin{acknowledgments}
We thank Ken Shiozaki for valuable comments and discussions.
This work is supported by JSPS KAKENHI Grant No. 22K03513.
\end{acknowledgments}

\appendix

\section{The variational argument with the Lieb-Schultz-Mattis operator}
\label{app:LSM}
We quickly review the Lieb-Schultz-Mattis argument in one dimension~\cite{Bohm1949,LSM1961,
Oshikawa1997,Nakamura2002,Tasaki2018,Tasaki2021}.
We consider the variational state $\ket{\Phi}=U_x\ket{\Psi}$ for the uniquely gapped ground state $\ket{\Psi}$.
The fermion creation operator behaves as $U_xc_j^{\dagger}U_x^{-1}=e^{i2\pi j/L}c_j^{\dagger}$
and a straightforward calculation gives an estimate for the energy difference between these two states,
\begin{align}
\braket{\Psi|U_x^{\dagger}H_0U_x|\Psi}-\braket{\Psi|H_0|\Psi}\leq \frac{C}{L},
\label{eq:Evar}
\end{align}
where $C$ is a positive constant which is independent of $L$.
From this inequality, one can obtain an inequality for the variational wavefunction as follows.
We expand the variational state in terms of the normalized energy eigenstates $\ket{\Psi_n}$ as
$\ket{\Phi}=z\ket{\Psi}+\sum_{n\geq 1}z_n\ket{\Psi_n}$, where $|z|^2+\sum_{n\geq1}|z_n|^2=1$ because
$U_x$ is a unitary operator.
Then, Eq.~\eqref{eq:Evar} is rewritten as 
\begin{align}
|z|^2E_0+\sum_{n\geq1}|z_n|^2E_n-E_0\leq \frac{C}{L},
\end{align}
where $E_n$ is the $n$-th eigenenergy.
We denote the energy gap between the ground state and the first excited state as $E_1-E_0=\Delta E>0$.
Clearly, $E_n-E_0\geq \Delta E$ for higher excited states.
The above inequality immediately implies that 
the deviation of the variational state from the ground state, $\ket{\delta\Psi}=\sum_{n\geq1}z_n\ket{\Psi_n}$,
is vanishing in the thermodynamic limit as
\begin{align}
\braket{\delta\Psi|\delta\Psi}=\sum_{n\geq 1}|z_n|^2\leq \frac{C}{L\cdot \Delta E}.
\end{align}
Equivalently, $|z|^2=1-C/(L\Delta E)$ and the variational state is nearly proportional to the ground state,
$\ket{\Phi}=U_x\ket{\Psi}=e^{i2\pi P_x}\ket{\Psi}+\ket{\delta\Psi'}$ with $\ket{\delta\Psi'}=(z-e^{i2\pi P_x})\ket{\Psi}
+\ket{\delta\Psi}$ and $2\pi P_x={\rm arg}z$.
This property is expressed as $\braket{\delta\Psi'|\delta\Psi'}=O(1/L)$ for simplicity and is used
in the main text and Appendix~\ref{app:BC}.

\section{Details of numerical calculations of Resta formula}
\label{app:Ux}
The Resta formula for the charge polarization can be easily evaluated for a non-interacting fermion system
with use of the single-particle wavefunctions.
Let $\{u_{jn}\}_{n=0}^{L-1}$ be the eigenvectors of the single-particle Hamiltonian,
where the eigenvalues are in the ascending-order.
Then the many-body ground state is given by
\begin{align}
\ket{\Psi}=\prod_{n=0}^{N-1}\gamma_n^{\dagger}\ket{0},
\end{align}
where $\gamma_n=\sum_j u^{\ast}_{jn}c_{j}$ are the single-particle modes diagonalizing the Hamiltonian.
Using $U_xc_j^{\dagger}U_x^{-1}=e^{i2\pi j/L}c_j^{\dagger}$,
we obtain
\begin{align}
\bra{\Psi}U_x\ket{\Psi}&=\bra{\Psi}\prod_n\left(U_x\gamma_n^{\dagger}U_x^{-1}\right)U_x\ket{0}\nonumber\\
&=\det {\mathcal U}_N\cdot e^{-2\pi i/L\sum_jj\nu},
\end{align}
where $({\mathcal U}_N)_{mn}=\sum_{j=0}^{L-1}u_{jm}^{\ast}u_{jn}e^{i2\pi j/L}, (m,n=0,1,\cdots N-1)$.

We remark on fluctuations of the polarization within the Resta formula.
One can introduce a parameter $\theta$ in the Resta formula as 
\begin{align}
P_x(\theta)&=\frac{1}{2\pi}{\rm Im}\log\bra{\Psi}U_x(\theta)\ket{\Psi},\\
U_x(\theta)&=\exp\left( \frac{i\theta}{L}\sum_j j(n_j-\nu)\right).
\end{align}
It is straightforward to calculate $\braket{U_x(\theta)}$ for general $\theta$ similarly to the above.
Then, higher order powers of the charge polarization can be obtained as
\begin{align}
P_x^{(n)}= \Bigg|\Braket{\frac{\partial^n U_x(\theta)}{\partial (i\theta)^n} \Big|_{\theta=0} }\Bigg|^{\frac{1}{n}}.
\label{eq:Pn}
\end{align}
We can numerically evaluate 
the higher order powers by polynomial fitting of $\braket{U_x(\theta)}$ as a function of $\theta$.
Under the periodic boundary condition,
we find that $P_x^{(2n+1)}<P_x^{(2n+3)}$ 
and $P_x^{(2n)}<P_x^{(2n+2)}$ respectively, and they all take 
different values.
Therefore, fluctuations of the charge polarization seem to exist in the present insulator, 
which may be simply due to the fact that
the operator $\hat{P}_x$ is not a good operator under the periodic boundary condition.
On the other hand, under the open boundary condition,
we see that $P_x^{(n)}\simeq |P_x(2\pi)|$ numerically holds for all $n=1,2,\cdots$.
Therefore, there are no fluctuations of the charge polarization in the present system
similary to standard thermodynamic quantities,
although the polarization itself is not a thermodynamic quantity.
We note that, in the numerical calculations, 
the fluctuations under the periodic boundary condition arise only from the inter unit cell 
contribution $\braket{U_{\rm inter}}$ and the intra unit cell contribution $\braket{U_{\rm intra}}$ 
does not have fluctuations, which is consistent with the general argument in the main text.
Under the open boundary condition, there are no fluctuations in both contributions.

\section{Robustness of the charge polarization to boundary conditions}
\label{app:BC}
One would naively expect that
the charge polarization $P_x={\rm arg}\bra{\Psi}U_x\ket{\Psi}/2\pi$ is robust to boundary conditions
in an insulator.
Indeed, we have numerically checked that $P_x$ of the present non-interacting model does not change
under the periodic boundary condition and the open boundary condition.
It was shown previously that the expectaion value of 
a local operator whose support is far way from the boundary is almost unchanged by boundary
conditions in a gapped ground state based on the exponential decay of correlation functions
~\cite{Watanabe2018}. 
However, the operator $U_x$ is non-local and 
some discussions may be necessary for the robustness of $\braket{U_x}$.

Here, we provide an argument based on the locality of the boundary condition and gapped nature of the
ground state by following the previous studies~\cite{Watanabe2018,Haegeman2013,Koma2015}, although it is not rigorous.
We introduce a parameter $\lambda\in[0,1]$ in the hopping integral on the bond $\braket{L-1,0}$
as $t_{L-1}\to \lambda t_{L-1}$, where $\lambda=0$ corresponds to the open boundary condition and
$\lambda=1$ to the periodic boundary condition.
It is assumed that there are no gapless edge states under the open boundary condition and
the system is uniquely gapped for all $\lambda\in[0,1]$.
(If there exists a gapless edge state at $\lambda=0$, one may introduce a small perturbation to gap out it and 
turn off the perturbation at the end.)
Then, the ground state is almost an eigenstate of $U_x$ and the variational state can be written as
$U_x\ket{\Psi(\lambda)}=e^{i2\pi P_x(\lambda)}\ket{\Psi(\lambda)}+\ket{\delta\Psi(\lambda)}$
with $||\ket{\delta\Psi(\lambda)}||^2=O(1/L)$ as mentioned in Appendix~\ref{app:LSM}. 
We have
\begin{align}
\frac{d}{d\lambda}\bra{\Psi}U_x\ket{\Psi}
&=\Braket{\frac{d\Psi}{d\lambda}|U_x|\Psi}
+\Braket{\Psi|U_x|\frac{d\Psi}{d\lambda}} \nonumber\\
&\simeq e^{i2\pi P_x}\frac{d}{d\lambda}\braket{\Psi|\Psi} \nonumber\\
&\quad +\Braket{\frac{d\Psi}{d\lambda}|\delta\Psi} + \Braket{\delta\Psi|\frac{d\Psi}{d\lambda}},
\end{align}
where we have neglected the small error from $|z-e^{i2\pi P_x}|$.
The first term in the above equation is zero.
On the other hand,
the inner product $\Braket{\frac{d\Psi}{d\lambda}|\delta\Psi}$ 
can be $O(1)$ for a general change of the Hamiltonian, although $||\ket{\delta \Psi}||^2=O(1/L)$.
For example, $\braket{U_x}$ can change when we vary the model parameters in the bulk region of the system
such as the Thouless pumping as discussed in the main text.
However, since the change of the Hamiltonian is local in the present setup,
we can claim that the inner product is exponentially small under an assumption. 
Here, we suppose that there exist local operators such that $|\braket{\Psi_n|O_n^{(0)}|\Psi}|\geq f_n>0$, 
where the distance between their supports $\Omega_n^{(0)}$ and the boundary bond $\braket{L-1,0}$ is $O(L)$.
Namely, the excited states above the gap can be derived from local perturbations on the ground state.
Then, it was shown that one can construct local operators $O_n$ whose supports $\Omega_n$ are finite and
include those of $O_n^{(0)}$ ~\cite{Watanabe2018,Haegeman2013,Koma2015}.
The excited states can be well approximated by these operators as $\ket{\Psi_n}=O_n\ket{\Psi}
/||O_n\ket{\Psi}||$,
where errors are exponentially small with respect to $R={\rm dist}(\partial \Omega_n,\Omega_n^{(0)})$.

Now, we write the Hamiltonian as $H(\lambda)=H(0)+\lambda W$ with 
$W=-t_{L-1}(c_{L-1}^{\dagger}c_0+c_0^{\dagger}c_{L-1})$.
By differentiating $H(\lambda)\ket{\Psi(\lambda)}=E_0(\lambda)\ket{\Psi(\lambda)}$, we obtain
\begin{align}
\Braket{\Psi_n|\frac{d\Psi}{d\lambda}}=\frac{1}{E_n-E_0}\braket{\Psi_n|\delta W|\Psi},
\end{align}
where $\delta W=W-\braket{\Psi|W|\Psi}$. Note that the energy difference is greater than the excitation gap,
$E_n-E_0\geq \Delta E$.
Then the inner product is rewritten as
\begin{align}
\Braket{\delta\Psi|\frac{d\Psi}{d\lambda}}
&\simeq \sum_{n\geq1}\frac{z_n}{E_n-E_0}\frac{\Braket{\Psi|O_n^{\dagger}\delta W|\Psi}}{||O_n\ket{\Psi}||}.
\end{align}
It is known that, for local operators $A, B$ with the distance $d$ between their supports,
the correlation function is exponentially small,  $|\braket{\Psi|\delta A\delta B|\Psi}|\leq {\rm const}\cdot
e^{-{\rm const}\cdot d}$ in a gapped system~\cite{Hastings2006}. 
From this exponential decay of correlation functions and $0=\braket{\Psi_n|\Psi}\simeq \braket{\Psi|O_n^{\dagger}|\Psi}
/||O_n\ket{\Psi}||$,
we conclude that the numerator in the above equation is exponentially small in the system size $L$
with an appropriately chosen $\Omega_n$ such that ${\rm dist}(\Omega_n,\braket{L-1,0})=O(L)$.
This implies that $\frac{d}{d\lambda}\braket{\Psi|U_x|\Psi}\simeq0$
and the charge polarization
within the Resta formula is independent of the parameter $\lambda$.
Especially, it does not change for the open boundary condition and periodic boundary condition, $P_x(0)=P_x(1)$.
Further, the above argument is applicable also to general perturbations 
localized around the edges and hence the charge polarization is robust to them.

Finally, one might think that the robustness of the charge polarization will immediately follow from the previous study
for local operators~\cite{Watanabe2018}, if one uses a local variant of the Lieb-Schultz-Mattis operator 
whose support is far away from the boundary bond~\cite{Tasaki2021}.
We naively expect that expectation values of the global Lieb-Schultz-Mattis operator and
the local one are equal under the periodic boundary condition.
However, we are interested in the physical quantity $P_{\rm edge}$ where the edge sites are included
in the operator support. 
In this case, the supports of the local Lieb-Schultz-Mattis operator and the boundary operator $W$
are not separated and the robustness does not follow immediately.

\section{Direct calculation of polarization for Eq.~\eqref{eq:Psi3}}
\label{app:psi3}
We can directly calculate the polarization in the ground state Eq.~\eqref{eq:Psi3} under the periodic boundary
condition.
Since $U_xc_j^{\dagger}U_x^{-1}=e^{i2\pi j/L}c_j^{\dagger}$, the dimers behave as
$U_x(c_j^{\dagger}+c_{j+1}^{\dagger})\ket{0}
=e^{i2\pi (j+1/2)/L}(e^{-i\pi/L}c_j^{\dagger}+e^{i\pi/L}c_{j+1}^{\dagger})U_x\ket{0}$.
Note that $|\braket{U_x}|\to1$ as $L\to\infty$,
because $\prod_{k=0}^{L_3-1} \cos(\pi/L)\to 1$.
The argument of the expectation value is 
\begin{align}
\frac{1}{2\pi}{\rm arg}\braket{U_x}
&=\frac{1}{L}\sum_{k=0}^{L_3-1}(3k+5/2)-\frac{1}{L}\sum_{j=0}^{L-1}j\nu
=\frac{1}{2}.
\end{align}
It is also straightforward to calculate $P_{\rm bulk}$ for the $k$-th unit cell (the sites $j=3k,3k+1,3k+2$),
\begin{align}
3P_{\rm bulk}&=\Braket{\sum_{l=0}^2(3k+l)(n_{3k+l}-1/3)} \nonumber \\
&=\frac{1}{4}[3k\cdot2+(3k+1)\cdot0+(3k+2)\cdot2]-(3k+1) \nonumber\\
&=0.
\end{align}
These results are fully consistent with the symmetry argument in the main text.

\bibliography{ref}

\end{document}